\documentclass[namedreferences,linksfromyear,optionalrh,solaromanenum]{spr-sola}
\usepackage{hyperref}
\hypersetup{
 colorlinks=true,
 citecolor=blue,
 linkcolor=blue,
 urlcolor=blue}

\usepackage{graphicx}                    
\usepackage{color}                       

\chardef\us=`\_

\begin{document}

\begin{frontmatter}

\title{A Statistical Study of $\delta$-Sunspots from Solar Cycle 23 to 25}

%
\author[addressref={aff1},corref,email={rlouis@prl.res.in}]{\inits{R.E.L}\fnm{Rohan Eugene}~\snm{Louis}\orcid{0000-0001-5963-8293}}

%
\runningauthor{R.E. Louis}
\runningtitle{A Statistical Study of $\delta$-Sunspots from Solar Cycle 23 to 25}

\address[id={aff1}]{Udaipur Solar Observatory, Physical Research Laboratory, Dewali Badi Road, Udaipur - 313001, Rajasthan India}

\begin{abstract}
Sunspots or active regions with a $\delta$-magnetic configuration are known to be associated with strong eruptions such as 
flares and mass ejections. 
This article, investigates the relationship between $\delta$-active regions and flares over the course of 
three solar cycles, from 1996 to 2024, with respect to the former's area, lifetime, latitudinal distribution, and the phase of its 
magnetic complexity. 
Solar cycle 25, while still in progress, has produced the least number of $\delta$-active regions in comparison 
to the previous two solar cycles, yet the number of M- and X-class flares exceed that of cycle 24 by 25\%. Flare occurrence is 
higher in C-, M-, and X-class events during the presence of the $\delta$-configuration in an active region which is seen in 
all three solar cycles. The total number of flares produced by $\delta$- and  non-$\delta$-active regions were 15875 and
17033, respectively across all three solar cycles. The latter are dominated by B- and C-class flares, while the number of M- and X-class 
flares across all three solar cycles was significantly less than compared to $\delta$-ARs.
The median lifetime of an active region in the $\delta$-phase is about five days while it is about eight days 
in the non-$\delta$-phase. 
The typical number of flares produced by a $\delta$-active region is 20, with maximum values ranging from 80--156 
for lifetimes between 6--13 days. 
However, about 30\% of $\delta$-active regions do not produce flares when their lifetimes are between 6--12 days. 
The latitudinal distribution of $\delta$-active regions across the northern and 
southern hemispheres is nearly symmetric on either side of the equator for solar cycles 23 and 24, peaking around 
$\pm$10$^\circ$--20$^\circ$. For solar cycles 23 and 24, about 30\% of the host $\delta$-active 
regions have an area exceeding the mean value over the above latitudinal belt while for solar cycle 25, there is a 
large scatter possibly due to the cycle still being in progress. It remains to be seen if the 
latter phase of solar cycle 25 will be as active as its earlier phase and whether the number of $\delta$-active regions emerging 
during that period scale with the total sunspot number.
\end{abstract}

%

\end{frontmatter}

%
\section{Introduction}
\label{intro} 
Sunspots or active regions (ARs) represent the largest and strongest manifestations of the solar magnetic field at the
photosphere. The presence of a magnetic field in sunspots is credited to George Hale \citep{1908ApJ....28..315H}, 
following the discovery of the Zeeman effect \citep{1897Natur..55..347Z}. 
Sunspots serve as excellent indicators of solar activity and it has been well established that solar eruptions such as flares and 
coronal mass ejections (CMEs) often originate in large ARs with a complex magnetic structure 
\citep{1987SoPh..113..267Z,1994SoPh..149..105S,2000ApJ...540..583S,2012SoPh..281..639L,2017ApJ...834..150Y}. These eruptions also 
exhibit a strong cyclic modulation with the sunspot number and area 
\citep[][and references therein]{2001A&A...375.1049T,2004SoPh..219..343J,2009ApJ...691.1222R,
2011JASTP..73..264Z,2012LRSP....9....3W,2015LRSP...12....4H,2015Ap&SS.356..215P} that varies quasi-periodically over a period of about 11 years.
The earliest classification of ARs based on their magnetic complexity was the Mount Wilson or Hale classification 
\citep{1919ApJ....49..153H} which comprised three major groups, namely, unipolar ($\alpha$), bipolar ($\beta$), and 
multipolar or mixed polarity ($\gamma$) ARs. The fourth class, namely the $\delta$-class, represents ARs that comprise
opposite polarity umbral cores that share a common penumbra \citep{1960AN....285..271K,1965AN....288..177K}.   
While an AR is generally classified as a combination of one or more of the above classes, the five main Hale classes of ARs
are $\alpha$, $\beta$, $\gamma$, $\beta\gamma$, and $\delta$ \citep{2015LRSP...12....1V}. In addition to this, 
the McIntosh classification \citep{1990SoPh..125..251M} uses white-light images to determine the complexity of an AR, while the 
Solar Optical Observing Network (SOON), jointly operated by the National Oceanic and Atmospheric Administration (NOAA) 
and the US Air Force \citep{2016SoPh..291.3123B}, uses a combination of H$\alpha$ images and line-of-sight magnetograms to 
categorize ARs into the following seven classes, namely, $\alpha$, $\beta$, $\gamma$, $\beta\gamma$, $\beta\delta$, $\beta\gamma\delta$, 
and $\gamma\delta$.

Several studies have been carried out to determine the role and association of complex sunspot groups, particularly those having a $\delta$-
configuration, with strong solar eruptions \citep[][and references therein]{2019LRSP...16....3T}. \cite{2002ApJ...569.1016F,2006ApJ...644.1258F}
found that strong transverse fields at the polarity inversion line (PIL), a large shear angle, and a steep magnetic gradient were closely 
related to the occurrence of CMEs, while \cite{2007ApJ...655L.117S} found the total unsigned flux next to the PIL as a proxy for the upper limit of 
the X-ray flare class. The initial brightenings preceding flares have been observed in small-scale structures at the PIL \citep{2010ApJ...719..403L}, that
can trigger a large-scale eruption \citep{2017ApJ...838..134B,2018ApJ...856...43B}. In addition, strong shear flows at the PIL have also been known to
be associated with flares \citep{1976SoPh...47..233H,2003A&A...412..541M} which are often seen with strong rotating ARs
\citep{2009SoPh..258..203M,2012ApJ...761...60V} and proper motion of sunspots \citep{2014A&A...562A.110L}. \cite{2000ApJ...544..540L} found that 
``magnetic tongues'', which are extended magnetic structures of opposite polarities straddling the PIL, were also seen in flaring ARs along with 
sheared coronal loops, sigmoids, and `J'-shaped flare ribbons \citep{2007SoPh..246..365G,2009ApJ...693L..27C,2014SoPh..289.2041M}.
\cite{2011A&A...534A..47C} found that complex ARs, such as those with a $\beta\delta$-configuration, produced more CMEs than simple unipolar ARs.
\cite{2014MNRAS.441.2208G} reported that as much as 88\% of X-class flares originated in $\beta\gamma\delta$-ARs during solar cycles (SCs) 22–-23. 
It has also been shown that the flare index and the time duration, for which an AR remains in 
the $\delta$-state, are positively correlated \citep{2015SoPh..290.2093T}. \cite{2016ApJ...820L..11J} found that only around 
16\% of all ARs are complex, based on data between 1992--2015, with the magnetic complexity varying with the SC. 
\cite{2019A&A...629A..45N} reported that the temporal evolution of $\alpha$- and $\beta$-ARs was different from that of $\beta\gamma$- and 
$\beta\gamma\delta$-ARs over the course of the SC with the former group following the sunspot cycle while the abundance 
of the latter peaked around two years later. \cite{2023NewA..10001972O} found that $\beta\gamma\delta$-ARs accounted for 70\% and 41\%
of X- and M-class flares, respectively between 1996--2018. 
\cite{2003ApJ...595.1277L, 2003ApJ...595.1296L} attempted to derive a set of unique parameters based on photospheric vector magnetograms
that could be used to distinguish flaring ARs from quiescent ones, which was further expanded by \cite{2015ApJ...798..135B}. However,
there is no single, ``smoking gun'' among the various AR characteristics described above that can be deemed responsible for producing flares.
Thus, in the context of magnetically complex ARs it is unclear, what fraction of $\delta$-ARs are quiescent, 
whether the number of $\delta$-ARs directly translates into enhanced flaring activity, and 
if the ratio of $\delta$-ARs to the total number of ARs is the same for all solar cycles.

In this article, I revisit the properties of $\delta$-ARs over three SCs from 1996--2024 by separating the $\delta$- as well as 
non-$\delta$-phases of the AR, and studying the number of flares, total flare intensity in X-rays, and the flare classes produced 
by these ARs. I also investigate the cyclic distribution of the various Hale classes that combine with a $\delta$-configuration, the 
area, lifetime, location of the host AR, and the dependence of the $\delta$-sunspot number with the flare productivity as a function of the SC. 
The rest of the article is organized as follows - the data sources and processing methods are elaborated in Sect.~\ref{data}, the results are 
described in Sect.~\ref{result}, while Sects.~\ref{discuss} and \ref{conclude} contain the discussion and concluding statements, respectively.


\section{Data Sources and Methods}
\label{data} 
\subsection{SolarMonitor}
\label{solmon}
The primary source of information on $\delta$-sunspots was extracted from SolarMonitor\footnote{\url{https://solarmonitor.org}} 
which comprises near-realtime and archived information on solar ARs and activity. In this article, I primarily used the 
summary table of the NOAA ARs located just beneath the thumbnail images. This table consists of  
seven columns that describe the NOAA AR number, the position of the AR, Hale Class, McIntosh Class, sunspot area expressed in millionths of the 
solar hemisphere (MSH), number of spots, and recent flares, respectively. Each row in the table corresponds to an AR which is also indicated
in the thumbnail images above. Most of the information in this table comes from the NOAA 
Space Weather Prediction Centre (SWPC)\footnote{\url{https://www.swpc.noaa.gov/products/solar-region-summary}} 
and is available for each date that can be accessed by parsing the date in Solar Monitor's uniform resource locator (URL). In order to extract 
the NOAA AR summary table, the source page for each date, available as a Hypertext Preprocessor (.php) file, was downloaded using a program written 
in Interactive Data Language (IDL) which would search for the signature HyperText Markup Language (HTML) tag corresponding to each column of 
the table. For instance, by counting the number of instances the keyword ``noaa\_number'' appeared as an HTML data table (enclosed within 
$<$td$><$/td$>$ tags) in the source file would indicate the total number of ARs identified by NOAA on that day. Omitting the McIntosh class 
all other columns were identified based on their signature keyword and stored as a string array. For extracting the flare information 
from the source page, the IDL program would search for all matches to the anchor tag ($<$a$><$/a$>$) for each NOAA number, remove the text 
ending with the corresponding flare's URL, and save the flare class, taking care to identify the current day's flare based on its HTML 
color coding, and ignore the previous day's flare information. Thus, the flare table would contain all the flares associated with each AR 
along with the date. These two tables were then concatenated into two final string arrays from 1996 January 01 to 2024 December 31. 
  
\subsection{Flare Catalog}
\label{flare}
The flare information from Solar Monitor was complemented with additional data from the Heliophysics Event 
Catalog (HEC)\footnote{\url{http://hec.helio-vo.eu/hec/hec_gui.php}} which contains the Geostationary Operational Environmental 
Satellite (GOES) X-ray flare list and is available as an extensible markup language (.xml) file for the chosen time interval. 
The flare information from this file was extracted in a similar way as described earlier with the Solar Monitor .php file excluding 
those table elements where there were no NOAA ARs. The total number of flares in the HEC and Solar Monitor catalogs were 30750 and 14736, 
respectively. It is to be noted that the latter does not record B-class flares. Removing duplicate entries from the two lists, the total 
number of flares spanning the time duration under study were 32908.
An additional flare catalog from the X-ray telescope \citep[XRT;][]{2007SoPh..243...63G} on board Hinode \citep{2007SoPh..243....3K}
was also used in this work \citep{2012SoPh..279..317W}. This catalog provides a list of flares from the time Hinode became operational
in November 2006 to the present day.

\subsection{Synoptic Sunspot Data}
\label{sidc}
The synoptic sunspot data comprising the daily sunspot number and the monthly mean sunspot number from the Solar Influences Data 
Center (SIDC)\footnote{\url{https://www.sidc.be}}, maintained by the Royal Observatory of Belgium, was used to cross-check the number 
of days where there was no data in the Solar Monitor catalog. 


\section{Results}
\label{result}

\subsection{Number of $\delta$-ARs \& Hale Class Association}
\label{delta_spot_no}

\begin{figure}[!h] 
\centerline{
\includegraphics[angle = 90,width=\textwidth]{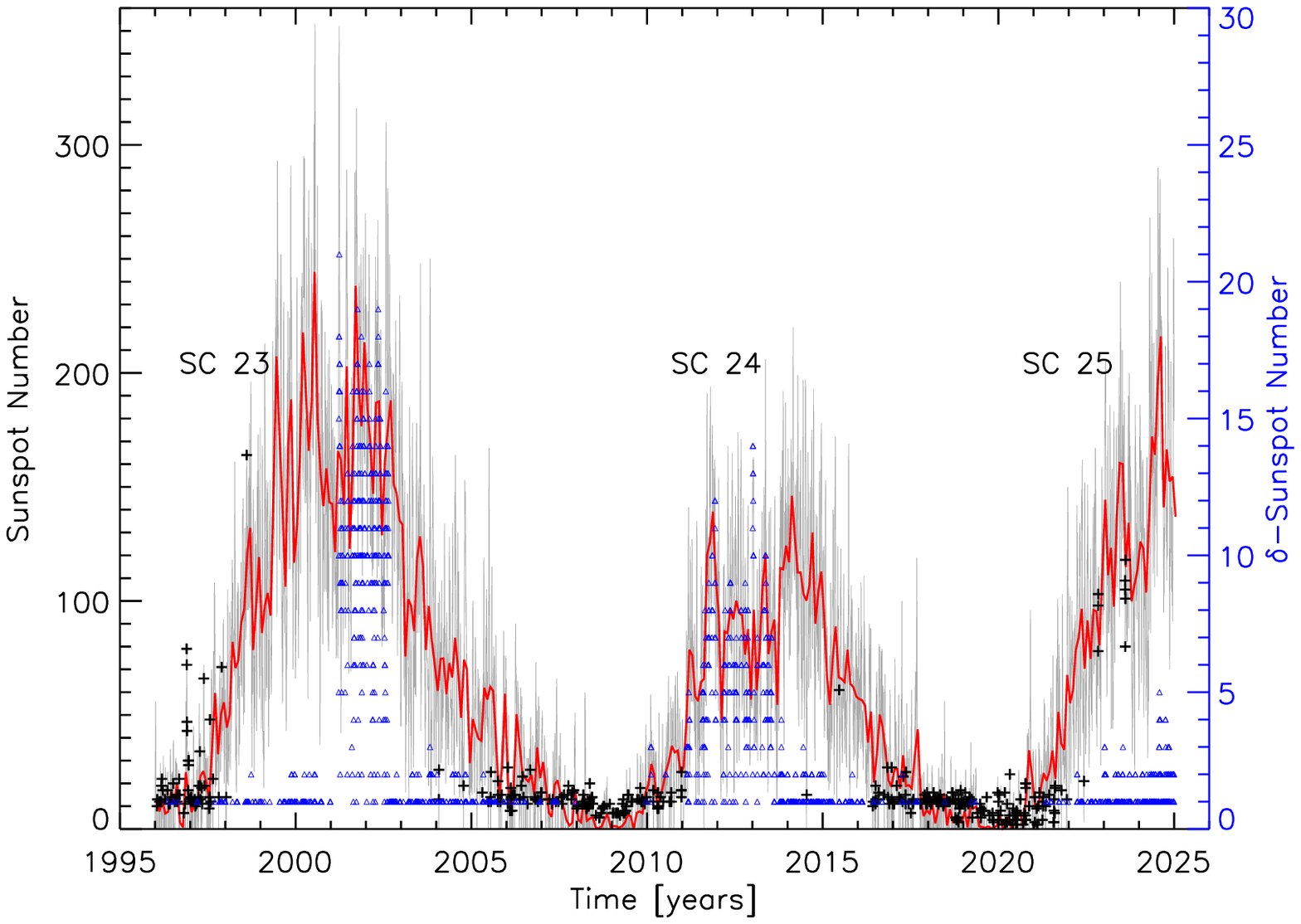}
}
\vspace{-10pt}
\caption{Sunspot numbers from 1996 January 01 to 2024 December 31. The grey and red lines correspond to the daily and 
monthly sunspot numbers from SIDC. The blue triangles indicate the daily number of $\delta$-sunspots recorded
in Solar Monitor. The black plus symbols represent the number of sunspots recorded by SIDC but not by Solar Monitor.}
\label{fig01}
\end{figure}

\begin{figure}[!h] 
\centerline{
\includegraphics[angle = 90,width=\textwidth]{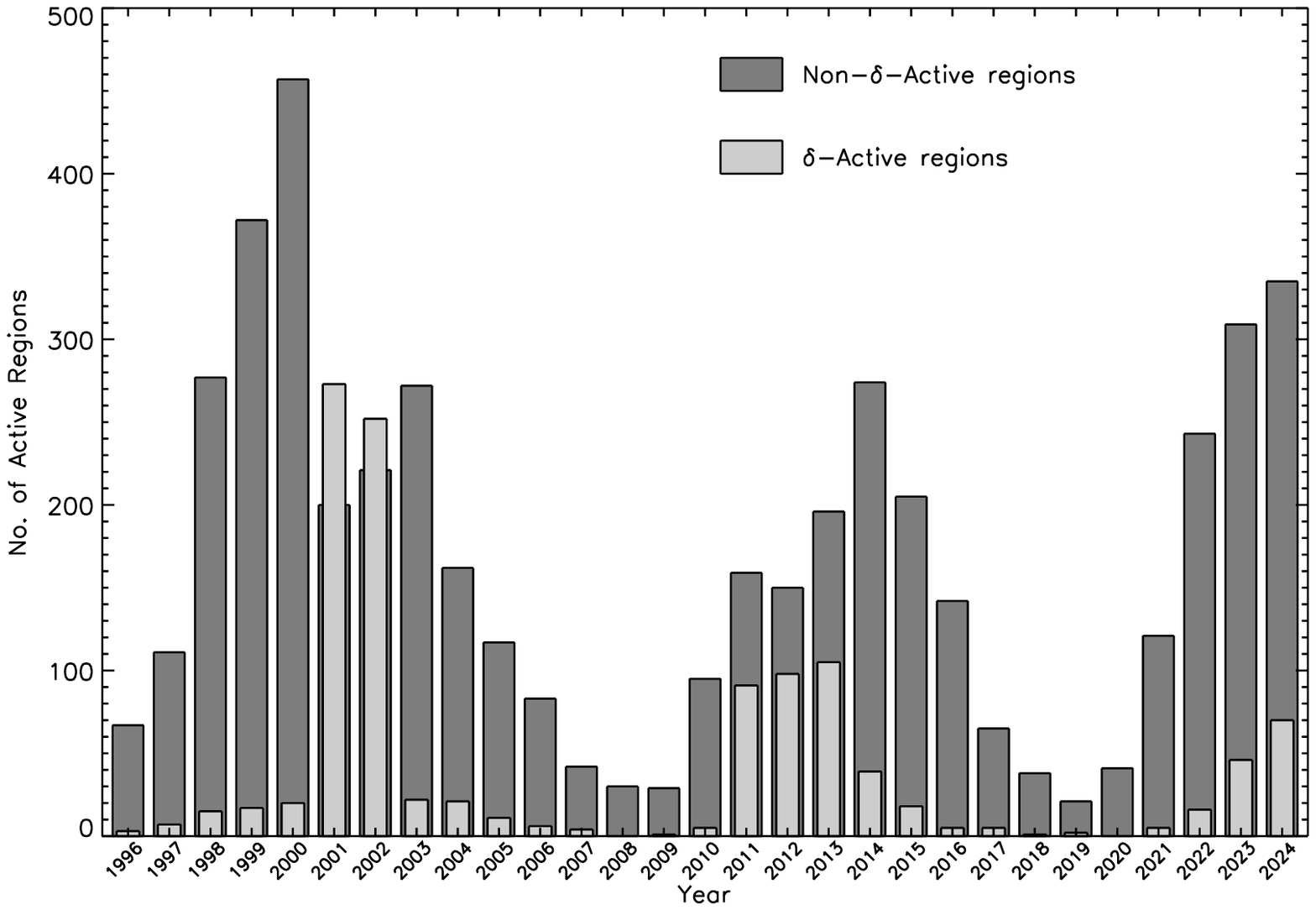}
}
\vspace{-10pt}
\caption{Yearly number of $\delta$- and non-$\delta$-ARs shown in light and dark grey, respectively from the SWPC dataset.}
\label{fig02}
\end{figure}

Figure~\ref{fig01} shows the daily sunspot number and the mean monthly sunspot number obtained from SIDC.
It is clear that SC 25 is stronger than SC 24 at nearly half of its duration and only about 65 sunspots
fewer than the maximum daily number in SC 23. The NOAA SWPC records of daily sunspot numbers 
do not have any data for about a year's length over the course of 29 years 
with the black plus symbols in the figure indicating the daily sunspot number from the SIDC catalog 
at those instances when the SWPC records were absent.
It is clear that these occasions, occur nearly exclusively, during the minimum phase of the SCs and in 
about 90\% of cases the number of daily spots not recorded by SWPC is less than 20. The absence of blue triangles 
indicates that $\delta$-sunspots were not observed/reported during the transition from one cycle to another for a period 
of two years between SC 23 to SC 24 and SC 24 to SC 25. The maximum number of $\delta$-spots
in a day varies from 21, 14, and five in the three cycles, respectively. It is also observed from Fig.~\ref{fig01}
that there appears to a be preferential period in the appearance of $\delta$-spots, where the increase in daily 
$\delta$-spot numbers, exceeding two spots a day, occurs around the second peak of SC 23 whereas in SC 24, the 
increase is seen during the first peak. It remains to be seen whether SC 25 will exhibit this temporal behavior
although it is to be noted that nearly half-way into the cycle the number of $\delta$-spots is still quite small
compared to the previous two cycles.

\begin{figure}[!h] 
\centerline{
\includegraphics[angle = 90,width=\textwidth]{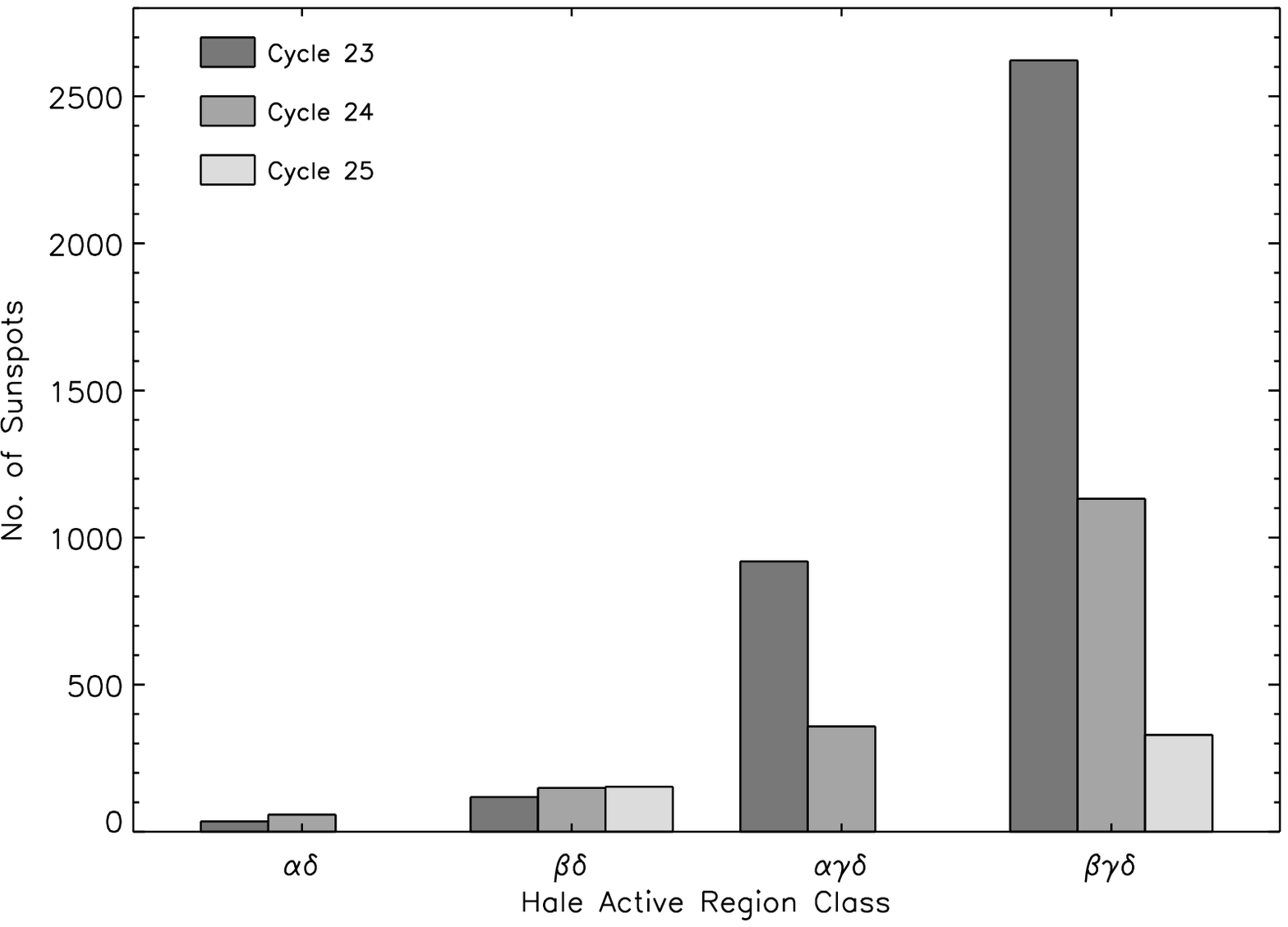}
}
\vspace{-10pt}
\caption{Distribution of various Hale classes of $\delta$-sunspots from SCs 23--25.}
\label{fig03}
\end{figure}

Figure~\ref{fig02} shows the yearly number of $\delta$- and non-$\delta$-ARs where the distribution is similar
to that seen in Fig.~\ref{fig01}. SC 23 shows the highest number of $\delta$-ARs exclusively during 2001 and 2002
with 273 and 252 ARs, respectively and exceeding the non-$\delta$-ARs by 73 and 31, respectively. The number of $\delta$-ARs
during the rest of SC 23 do not exceed 25 in comparison. This excess number of $\delta$-ARs during 2001 and 2002,
as seen in Fig.~\ref{fig01},
is most likely an artifact of the classification scheme used at SWPC/SolarMonitor. In Cycle 24, the number of $\delta$-ARs are highest between 
2011--2013 numbering 91, 98, and 105, respectively. Unlike Cycle 23, the number of $\delta$-ARs does not exceed the 
number of non-$\delta$-ARs. It remains to be seen if Cycle 25 will produce a greater number of $\delta$-ARs within the next 
couple of years in excess of 70 as of the end of 2024. Combining all three SCs, the total number of $\delta$- and non-$\delta$-ARs  
were 1158 and 4834, respectively.

\begin{figure} 
\centerline{
\includegraphics[angle = 0,width=\textwidth]{latitudinal-area-variation-delta-spots.eps}
}
\vspace{-10pt}
\caption{Latitudinal distribution of $\delta$-ARs for SCs 23--25. The number of ARs corresponding to the histogram are scaled
with respect to the y-axis on the left. The red solid line with triangles corresponds to the maximum area of the host AR in that 
latitudinal bin while the red dashed line represents the mean. The mean value has been scaled by a factor of two for clarity. 
Both quantities are scaled with respect to the red y-axis on the right. The blue crosses correspond to the fraction of $\delta$-ARs whose 
area exceeds the mean value in each bin and its scale is the same as the histogram y-axis on the left. The dotted horizontal line 
corresponds to a value of 30\% for this fraction. The numbers in the top right corner indicate the number of $\delta$-ARs in that cycle.}
\label{fig04}
\end{figure}

Figure~\ref{fig03} shows the Hale classification of $\delta$-spots across the three cycles with all the possible combinations 
appearing with the $\delta$-configuration since past studies focus on the $\beta\gamma$- and $\beta\gamma\delta$-types as the defining 
criteria for magnetically complex  ARs. The most commonly
occurring type are $\beta\gamma\delta$-spots followed by the $\alpha\gamma\delta$-class. Both these categories reflect the 
cyclic decrease in the number of $\delta$-spots wherein SC 25 did not comprise any $\alpha\gamma\delta$-class as 
recorded till the end of 2024. These classes account for about 69\% and 22\% of all $\delta$-spot classes, respectively
over three SCs. The remaining 9\% of $\delta$-spots belong to the $\beta\delta$- and $\alpha\delta$-classes.  


\begin{figure}[!h] 
\centerline{
\includegraphics[angle = 90,width=\textwidth]{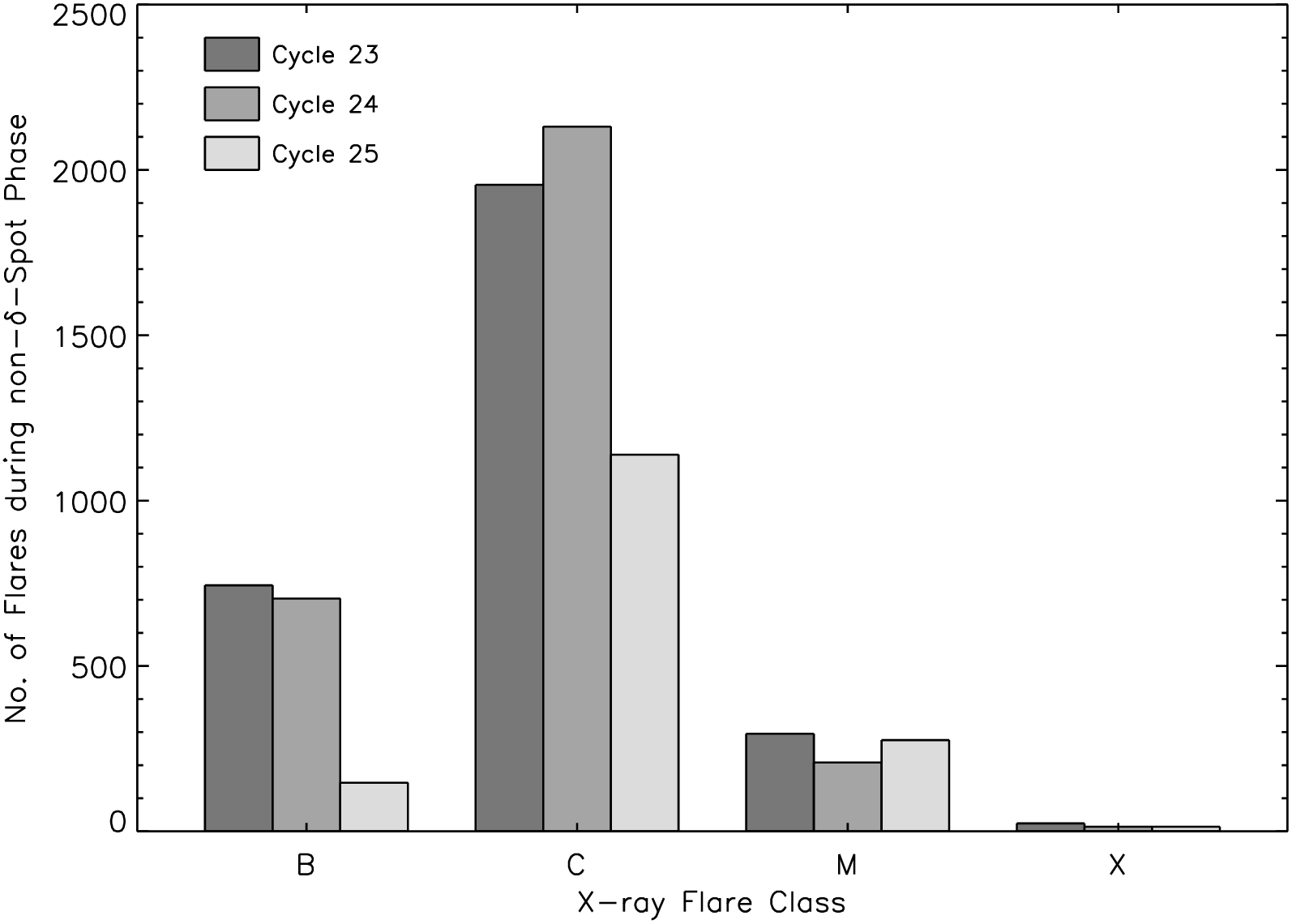}
}
\centerline{
\includegraphics[angle = 90,width=\textwidth]{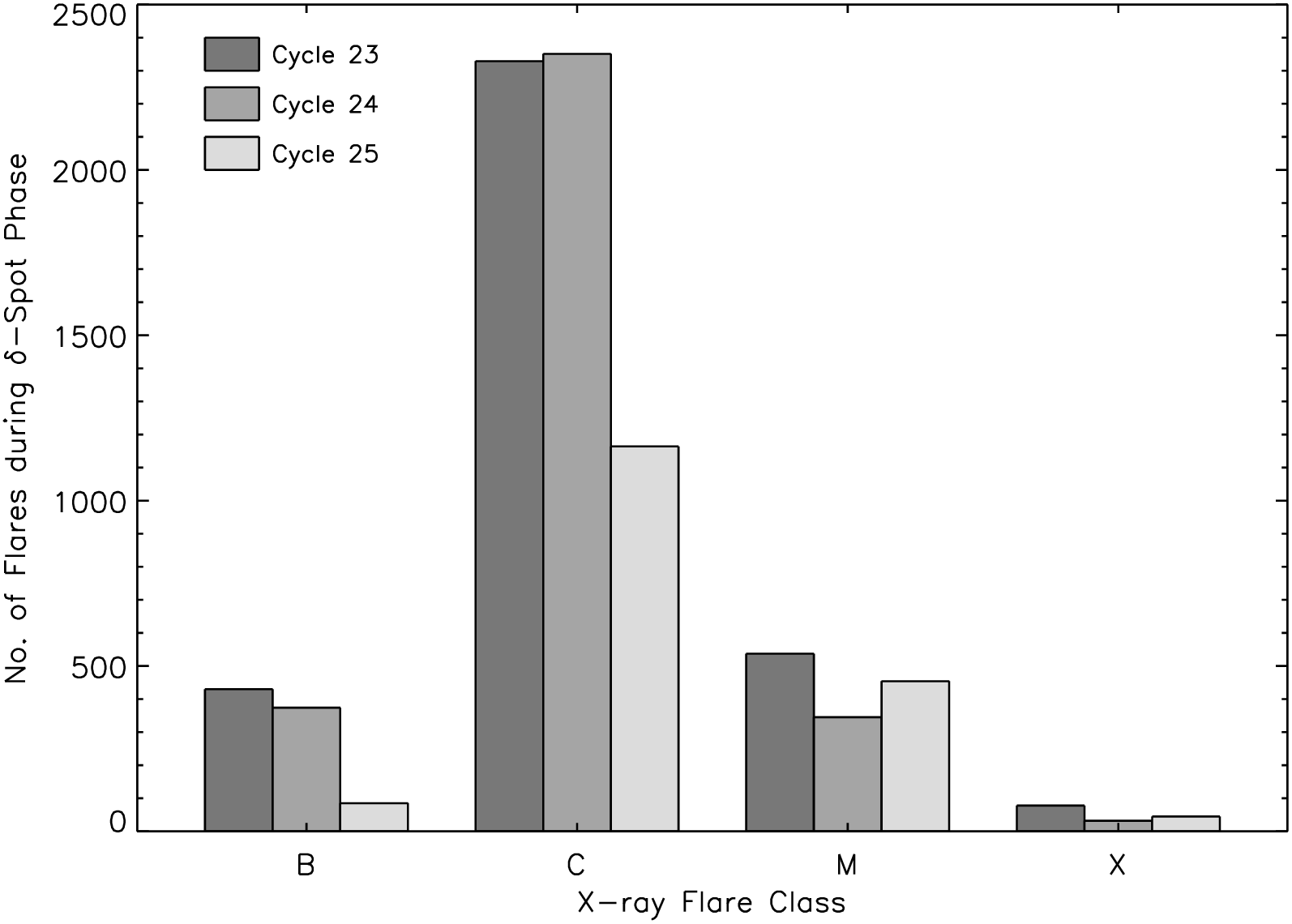}
}
\vspace{-10pt}
\caption{Distribution of flares in various X-ray classes over the three SCs during the non-$\delta$-phase (top panel)
and the $\delta$-phase (bottom panel) of an AR.}
\label{fig05}
\end{figure}

\subsection{Latitudinal Distribution of $\delta$-ARs and Area of Host AR}
\label{lat_delta_spot}
Figure~\ref{app-fig01} in the Appendix shows the butterfly diagram of $\delta$-ARs for the three solar cycles, 
where the colored plus symbols correspond to ARs having an area less than 1000 MSH, while the circles 
depict $\delta$-ARs with an area greater than 1000\,MSH, with the radius representing the area 
normalized by a value of 3500\,MSH. It is clear that the clutter of ARs between 2001 to the end of 2002 is a result of the over-counting at SWPC as
seen previously in Fig.~\ref{fig02} as well.
The information in the figure is summarized in Fig.~\ref{fig04} which shows the latitudinal distribution of the 
$\delta$-ARs as a histogram in $5^{\circ}$ bins along with the maximum and mean area of the host ARs in each latitudinal bin. The mean
values have been scaled by a factor of two for better visibility. 
While $\delta$-ARs are confined to the well-known sunspot activity belt of $\pm35^{\circ}$, the peak of the distribution
in general is symmetric and lies between 10$^\circ$--20$^{\circ}$ in both hemispheres for SCs 23 and 24. The maximum area of the host AR also
follows the peak of the latitudinal histogram, except that in SC 23 the values are nearly similar in both hemispheres at around 2500\,MSH, while
there is a clear hemispherical asymmetry in SC 24. This asymmetry, however, has to be understood in terms of the distribution of the mean 
value of the area (red dashed line) as well as the fraction of $\delta$-ARs whose area exceeds the mean value (blue crosses).
In the figure, it is evident that while the mean values of the host AR area exhibit a latitudinal trend they do not follow their maximum
counterparts and are confined to values of about 425\,MSH.
In addition, the dotted horizontal line shows that about 30\% of the $\delta$-ARs in each bin have an area exceeding the mean value, which is 
consistent for the active latitudinal belt as well as both cycles 23 and 24. The two outlier points in SC 23 at $30^\circ$ and
$35^\circ$ are related to the low number of $\delta$-ARs at that latitude. Taking both these aspects into account, I find that 
while the peak area in SC 23 at $-15^\circ$ is around 2750\,MSH, there are only two ARs whose area exceeds the maximum value of 1540\,MSH 
in the Northern Hemisphere at $15^\circ$. In comparison to SCs 23 and 24, cycle 25 exhibits a flatter latitudinal distribution, 
possibly due to the reduced number of 139 $\delta$-ARs recorded till the end of 2024. While the peak area values in the Southern Hemisphere 
are similar to that seen in the previous two cycles, there is a larger spread around the 30\% line denoting the fraction of $\delta$-ARs 
whose area exceeds the mean value in that latitudinal bin. One could attribute this characteristic difference in SC 25, with respect 
to the previous two cycles, on the fewer number of $\delta$-ARs for a solar cycle that is still in progression.     
The asymmetry in the maximum area between the northern and southern hemisphere however is due to a single AR with an area greater than 
the peak value of 1450\,MSH in the northern hemisphere, very similar to what was described earlier for SC 24.


\subsection{Flare Classes during $\delta$- and Non-$\delta$-Phase of ARs}
\label{flare_class}
In the following two sections, I shall describe the flare-related statistics during an AR's $\delta$- and 
non-$\delta$-phase. The $\delta$-phase corresponds to the instance when an AR was classified as having a $\delta$-type
magnetic configuration. This implies that during the transit of the AR there were periods when it 
possessed a $\delta$-type configuration and periods when it lacked one. 
Figure~\ref{fig05} shows the number of flares in various X-ray classes over the three
SCs during the non-$\delta$-phase (top panel) and the $\delta$-phase (bottom panel). It is evident that
C-class flares dominate in number over all cycles and in both phases by around 620 events. While the number of B-class
flares are reduced in the $\delta$-phase by around 700 events across all cycles, the M-class and X-class events during
the $\delta$-phase exceed the non-$\delta$-phase by around 555 and 100 events, respectively. It is also observed
that despite SC 25 showing fewer $\delta$-spots in comparison to SC 24, the former is associated with a 
higher number of M- and X-class flares in both phases of the AR, with 14 X-class flares in both cycles during the 
non-$\delta$-phase. 

The total number of flares across all three cycles in both phases of the AR amount to 15875 with 
the non-$\delta$-phase and $\delta$-phase accounting for 8224 and 7651 flares, respectively. 
The total number of flares combining the two phases in the three cycles are 6392, 6159, and 3324, respectively.
In comparison the number of flares in non-$\delta$ ARs across all SCs were 17033 which were dominated by 
B- and C-class flares, while the number of M- and X-class flares across all three SCs was significantly less than 
compared to $\delta$-ARs.


\begin{figure}[!h] 
\centerline{
\includegraphics[angle = 90,width=\textwidth]{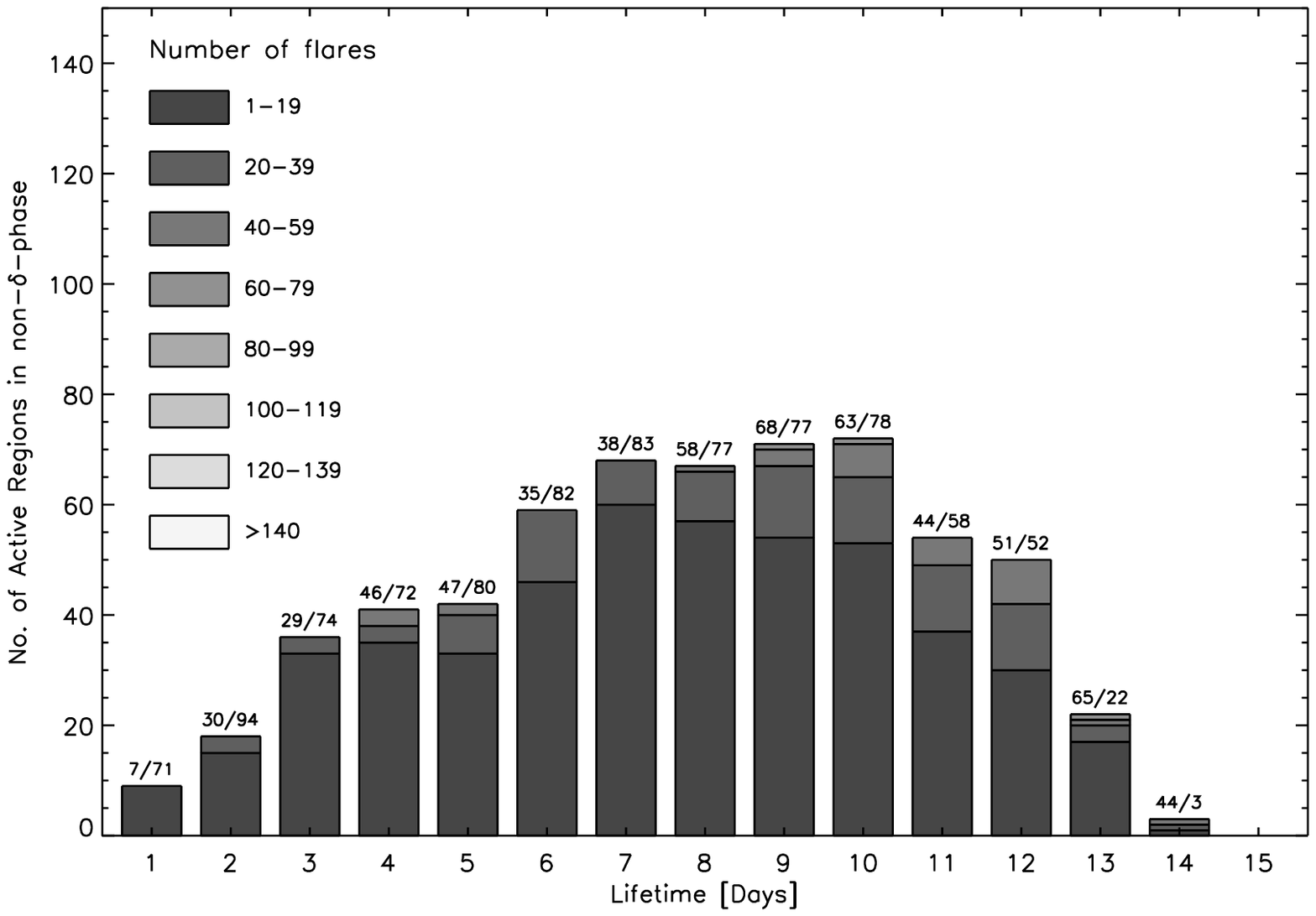}
}
\vspace{-10pt}
\centerline{
\includegraphics[angle = 90,width=\textwidth]{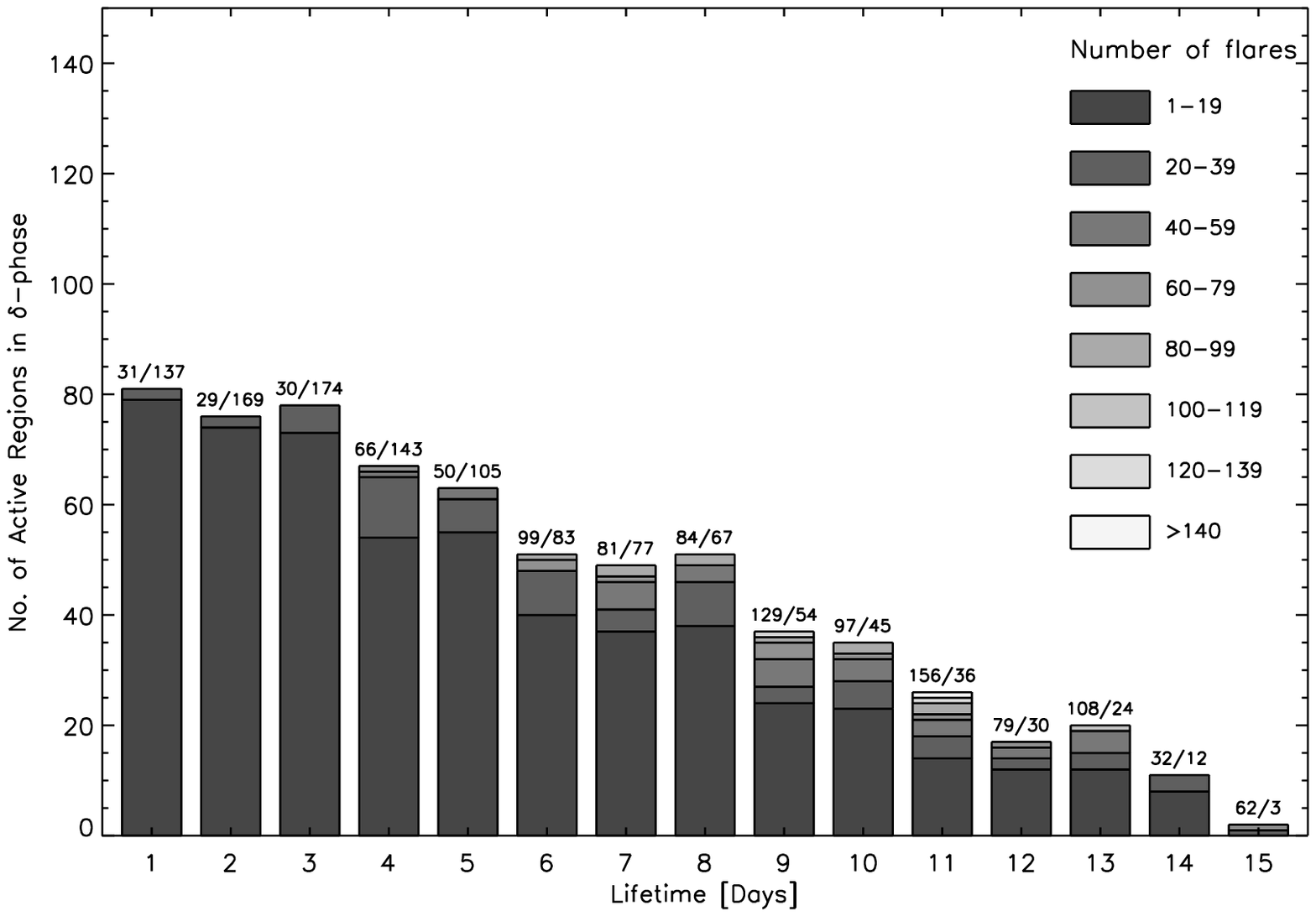}
}
\vspace{-10pt}
\caption{Histogram of flare producing $\delta$-ARs as a function of their lifetime during the 
two phases shown in the top and bottom panels, respectively. The color shades in each lifetime bin 
refer to the number of flares produced in intervals of 20 events as indicated in the legend.
See the text for an explanation on the numbers at the top of each histogram column.}
\label{fig06}
\end{figure}

\begin{figure}[!h] 
\centerline{
\includegraphics[angle = 90,width=\textwidth]{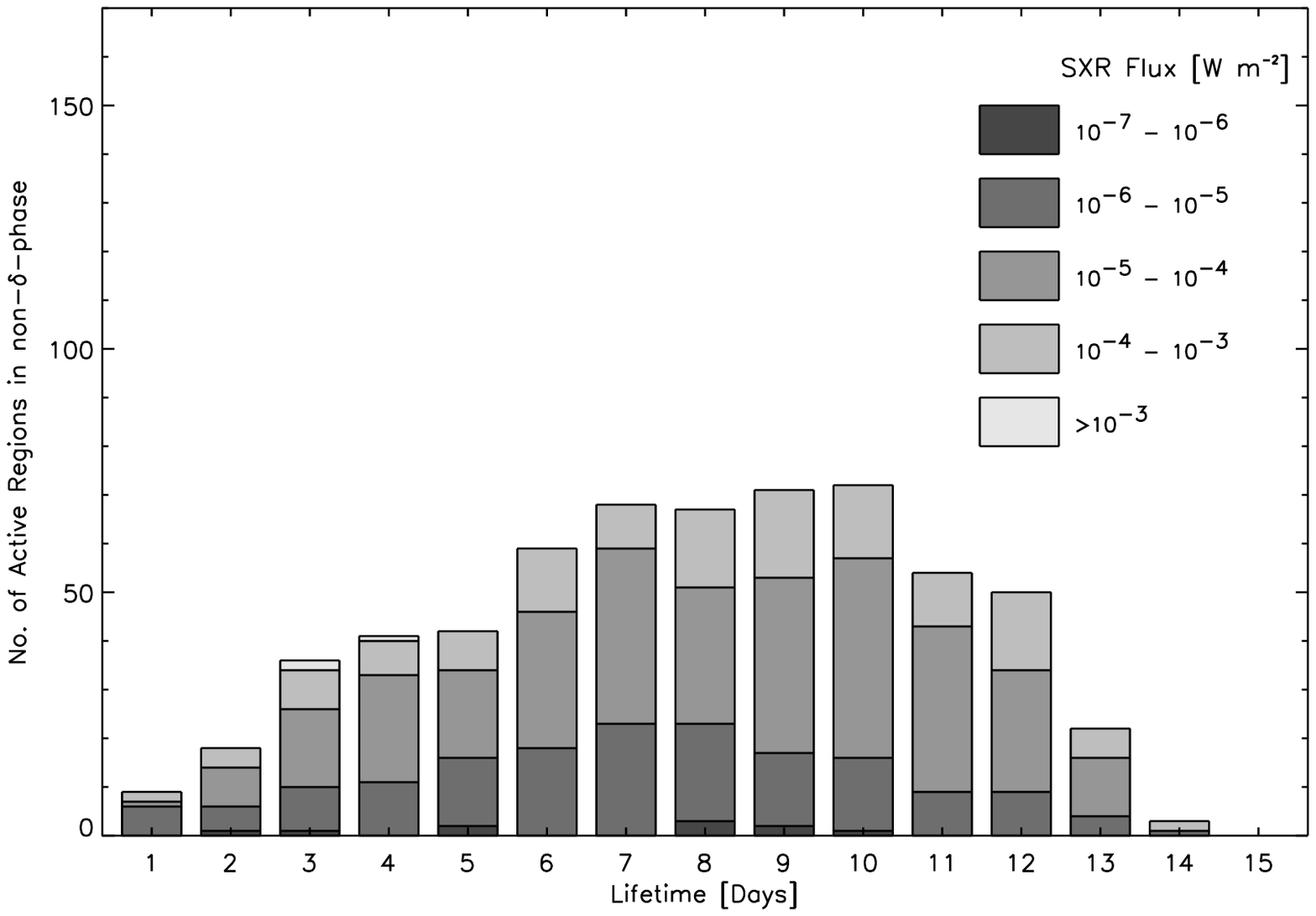}
}
\vspace{-10pt}
\centerline{
\includegraphics[angle = 90,width=\textwidth]{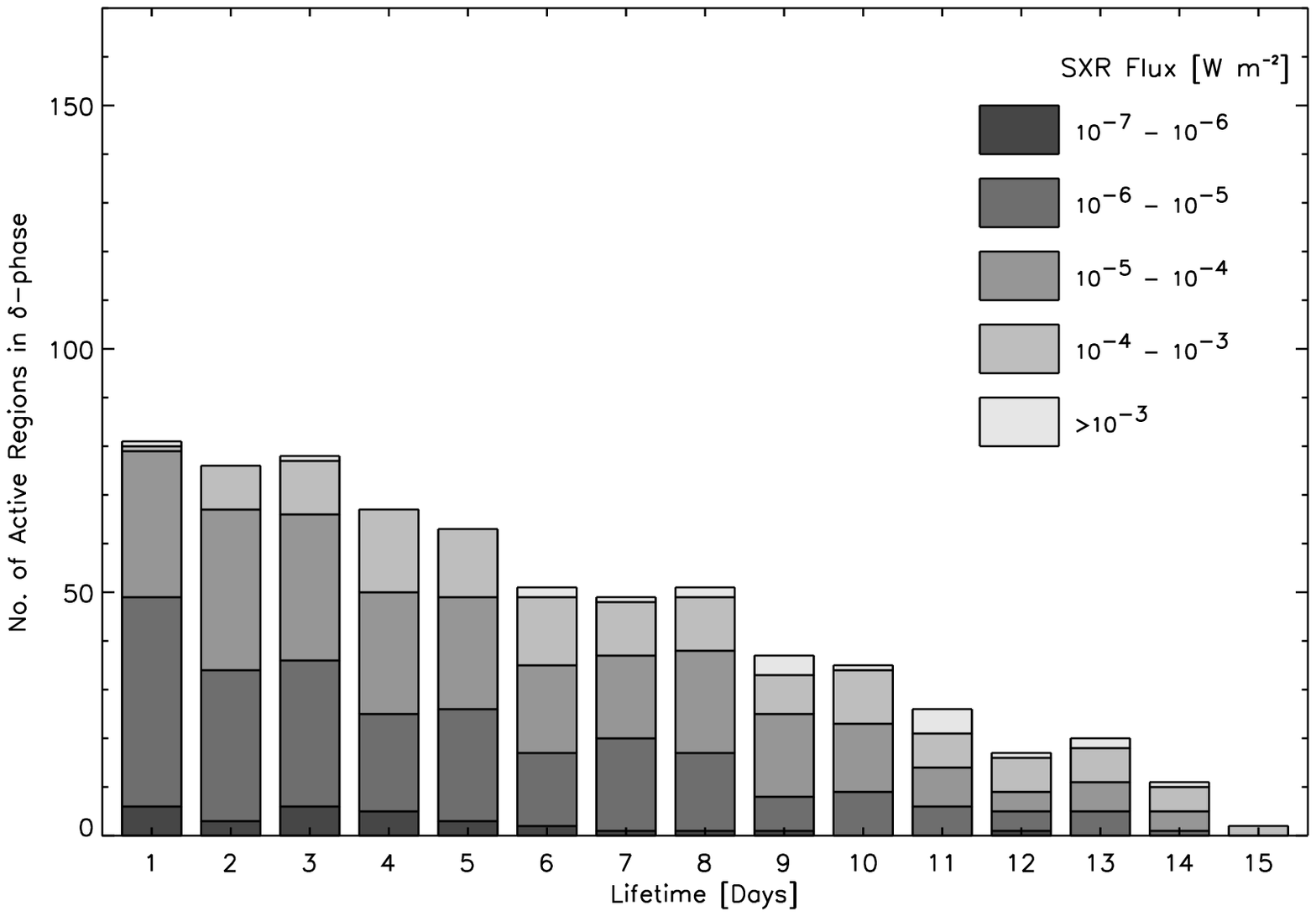}
}
\vspace{-10pt}
\caption{Same as Fig.~\ref{fig06} but for the flare intensity expressed in W\,m$^{-2}$ during the non-$\delta$ (top) 
and $\delta$-phases (bottom). The different shades of gray represent a step of an order of magnitude in intensity as shown in the legend.}
\label{fig07}
\end{figure}

\subsection{Flare Distribution in $\delta$-ARs as a Function of Lifetime}
\label{flare_lifetime}
The distribution of the flare number with the lifetime of $\delta$-ARs is shown in Figure~\ref{fig06}
during the two phases in the top and bottom panels, respectively. The histogram was constructed as a function of the lifetime
of the AR in the two phases separately and the color shades in each lifetime bin refer to the number of flares produced, across all flare
classes, in intervals of 20 events while the height corresponds to the number of ARs. 
The top of each bar comprises two numbers, with the first on 
the left representing the maximum number of flares produced by an AR, with the second showing the total number of ARs in that 
lifetime bin. In the non-$\delta$-phase, the lifetime distribution peaks around 7--10 days, with 82\% and above ARs, having 
lifetimes between 7--12 days, being flare productive. While the maximum number of flares produced by an AR in its non-$\delta$-phase
is 68, it is observed that the majority of the ARs (above 70\%) produce less than 20 flares over all lifetime bins as 
indicated by the darkest shade of grey in the figure. During the non-$\delta$-phase, ARs that produce more than 40 flares 
usually have lifetimes greater than eight days. The lifetime distribution of ARs in their $\delta$-phase, in comparison to the non-$\delta$-phase, 
shows a monotonic decrease with a majority of them having shorter lifetimes. However, the maximum number of flares produced
during the $\delta$-phase is greater than during the non-$\delta$-phase particularly for ARs with lifetimes greater than six days with
the difference reaching as large as 112 for a lifetime of 11 days. For lifetimes between 6--10 days the minimum and maximum difference
in the number of maximum flares are 26 and 64, respectively.  The figure also shows that during the $\delta$-phase, one to two ARs produce 
flares in excess of 80 events having lifetimes between 6--11 days with one AR producing more than 120 events when being in the $\delta$-
configuration for 11 days. One similarity with the non-$\delta$ phase is that more than 75\% of the ARs produce less than 20 flares.
 
Figure~\ref{fig07} shows a similar histogram as Figure~\ref{fig06} but for the total X-ray flux from the flares produced by each AR for
a specific lifetime bin during the non-$\delta$- and $\delta$-phase respectively, as indicated by the various shades of gray that represent
a step size of an order of magnitude in W\,m$^{-2}$. Since the sorting of the ARs is done on the basis of the lifetime, the distributions 
in the two phases are identical to Fig.~\ref{fig06}. It is seen that during the non-$\delta$-phase the total X-ray flux emitted during the 
flares lies in the $10^{-5}$--$10^{-4}$\,W\,m$^{-2}$, that is about 50\% or more for lifetimes in the 7--12 day range, with the exception 
of around 42\% for a lifetime of eight days. The next higher
energy band accounts for about 13\% to 32\% in the above lifetime range. In the $\delta$-phase, the energy ranges of 
$10^{-5}$--$10^{-4}$\,W\,m$^{-2}$ and $10^{-4}$--$10^{-3}$\,W\,m$^{-2}$ account for about 37\% and 22\%, respectively, 
of the distribution for lifetimes in the 3--8 day range. However, there is also a contribution from the highest energy 
range exceeding $10^{-3}$\,W\,m$^{-2}$ for lifetimes greater than six days, which is not observed in ARs during their 
non-$\delta$-phase. The fraction of the ARs associated with the highest energy range is about 11\% and 
19\% for lifetimes of nine and 11 days, respectively, which also coincide with the excess number of flares 
produced between the two phases of the AR.


\begin{figure}[!h]
\centerline{
\includegraphics[angle = 0,width=0.98\textwidth]{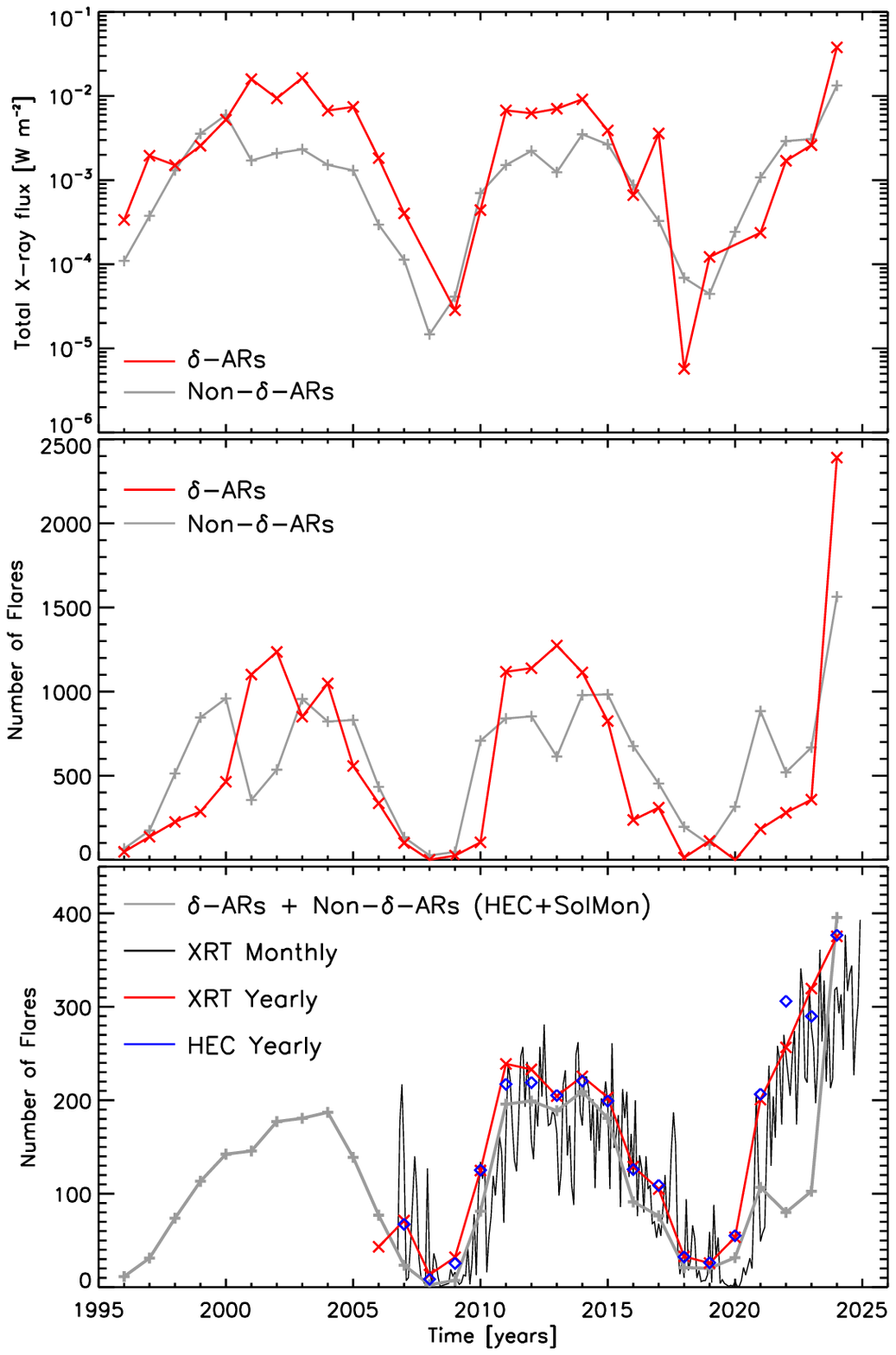}
}
\vspace{-5pt}
\caption{Variation of the number of flares and flare intensity as a function of time. Top panel: Total X-ray flare intensity
from $\delta$- and non-$\delta$-ARs indicated by red and gray colors, respectively and averaged over a year. Middle panel: Same as above
but for the number of flares. Bottom panel: The black and red lines represent the monthly and yearly number of flares, respectively 
from the Hinode XRT catalog. The gray line corresponds to the yearly number of flares from $\delta$- and 
non-$\delta$-ARs analyzed in this work. The blue diamonds represent the yearly number of all flares from the HEC alone. 
The yearly number of flares from all data sources has been reduced by a factor of ten.}
\label{fig08}
\end{figure}

\begin{figure}[!h]
\centerline{
\includegraphics[angle = 0,width=\textwidth]{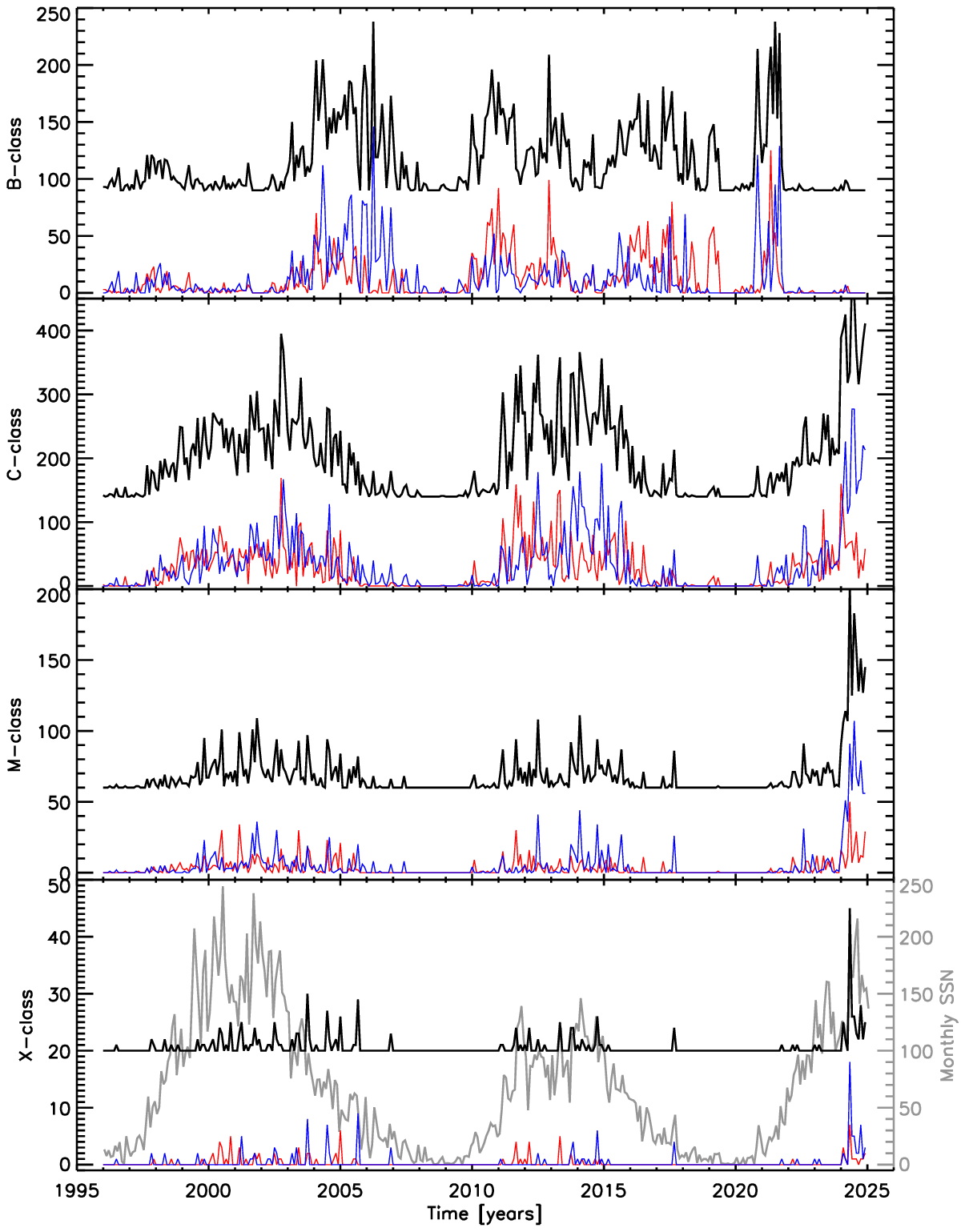}
}
\vspace{-25pt}
\caption{Variation of the monthly number of flares in different classes with time. The red and blue colors correspond to flares in the northern 
and southern hemispheres, respectively while the black line is the total number of flares shifted vertically for better visibility. The gray 
line in the bottom panel is the monthly mean sunspot number from SIDC.}
\label{fig09}
\end{figure}

\subsection{Flare Statistics of $\delta$- and Non-$\delta$-ARs}
\label{flares_all}
I finally compare the flares produced across all X-ray classes for $\delta$- and non-$\delta$-ARs as indicated in 
Figure~\ref{fig08}. The flares for the $\delta$-ARs include both phases of the $\delta$- and non-$\delta$-configuration
with the number of flares and the flare intensity summed over each respective AR for a one-year interval. During 
SC 23 the number of flares produced by $\delta$-ARs exceeded that by
non-$\delta$-ARs by a factor of 3.1 and 2.3 between 2001 and 2002, respectively. In SC 24, this factor was around 1.3, 1.3, and 2.1
between 2011, 2012, and 2013, respectively. The corresponding increase in the flare intensity in cycle 23 was 9.3 and 4.5, respectively 
while in cycle 24 it was around 4.5, 2.8, and 5.7, respectively for the same duration around the maximum phase of the cycle.
Despite hosting the fewest number of $\delta$-ARs, cycle 25 has produced 2391 flares in 2024 alone, accounting for 72\% of all flares
since 2019. This exceeds the maximum values in the previous two cycles by a factor of 1.9, while the number of flares from 
non-$\delta$-ARs in 2024 was 1564 and exceeded the maximum values in the previous two cycles by a factor of 1.6. In a similar way,
the flare intensity in 2024 from $\delta$-ARs exceeded the maximum values in SCs 23 and 24 by a factor of 2.3 and 4.1, respectively.

In order to verify that the conspicuous increase in the number of flares in SC 25 was not erroneous, the Hinode XRT flare catalog 
was utilized to cross-check the number of flares recorded independently between 2006--2024, that covers SCs 24 and 25. The 
bottom panel of Fig.~\ref{fig08} shows the total monthly number of flares in black while the red line is the total yearly number reduced
by a factor of ten to fit in the same scale as the former. The total number of flares used in this study combining both $\delta$- and 
non-$\delta$ ARs is shown in red, which is also reduced by a factor of ten and is essentially the sum of the gray and red plots from the 
middle panel of the figure. The yearly number of flares from both data sets is in good agreement and the apparent increase in the number of
flares in SC 25 particularly in the year 2024 is also consistent. There is a mismatch between the two data sets from 2021--2023 which 
is due to A-, B-, and weak C-class flares that occurred in the QS which were not associated with any AR. A large majority of these flares 
during this period are listed in the HEC but do not contain an AR number. This is illustrated by the blue diamonds which represent all flares in the HEC
that match the yearly number of events from the Hinode XRT catalog. Nevertheless,  
it is clear that SC 25 has been more flare productive than SC 24 as seen from both the monthly and yearly numbers.

For completeness, Fig.~\ref{fig09} shows the monthly number of flares of different classes with time for the two hemispheres separately 
(red and blue lines) as well as the total (black line). It is clear that the number of flares is dominated by the C-class, whose temporal 
variation matches that of the SC very closely. The number of B-class flares follows next but they do not have a preferred epoch with the 
SC as they can occur during i) the descending phase (SC 23, SC 24), ii) the main activity phase (SC 24), and iii) the ascending phase (SC 25). 
The M- and X-class flares are much fewer in number, but the sharp increase in the number of flares in SC 25, as seen in Fig.~\ref{fig08}, 
is reflected in all flare classes except B-class flares. It is also observed that the C-, M-, and X- flares in SC 25 predominantly occurred 
in the southern hemisphere.


\section{Discussion}
\label{discuss}
The magnetic complexity of solar ARs is the primary source of eruptions from the Sun which is a key component in driving and
influencing space weather. The source regions that produce strong flares and mass ejections often exhibit a 
$\beta\gamma$ and $\beta\gamma\delta$ configuration \citep{1983SoPh...88..275N,1987SoPh..113..267Z,1993A&A...272..609M,
1994SoPh..149..105S,1998ApJ...502L.181A,2000JASTP..62.1437V,2014MNRAS.441.2208G,2017ApJ...834...56T,2023ApJ...958....1L}. 
The latter in particular represents the most complex magnetic configuration \citep{2015ApJ...806...79F}
as it comprises a polarity inversion line within a sunspot as a consequence of two umbral cores
of opposite polarity residing within a common penumbra. This article investigates the properties of $\delta$-ARs from SC 23 to the 
present SC 25 to ascertain if the flare productivity during the presence of the $\delta$-type configuration were distinct from those 
during the non-$\delta$-type configuration.

$\delta$-ARs, in combination with all other Hale configurations, comprise about 20\% of all ARs from a total of 5992 ARs covering 
SCs 23 to 25, and is in agreement with the value of 16\% by \cite{2016ApJ...820L..11J}. SC 25, in particular, has produced the fewest 
number of $\delta$-ARs numbering 139 till the end of 2024, while SCs 23 and 24 comprised 651 and 368 $\delta$-ARs, respectively. 
The $\delta$-phase of an AR produces more C-, M-, and X-class flares than during the non-$\delta$-phase across all three SCs, while the 
opposite is true for B-class flares. Despite SC 25 showing the fewest number of $\delta$-ARs, the latter has produced more M- and X-class 
flares than their counterparts in SC 24. In addition, it has been observed that the total number of flares produced by $\delta$- and 
non-$\delta$-ARs in 2024 alone exceed the maximum values in the previous cycle which account for nearly 70\% of all flares in SC 25,
that include, C-, M-, and X-class flares. This result from the data set, combining SWPC and HEC resources, used in the present study is 
consistent with that from the Hinode XRT flare catalog as well. While SC 25 has shown a higher number of sunspots than the previous cycle,
the fact that a fewer number of $\delta$-ARs have produced more flares than cycle 24 would imply that at least upto this present juncture
of time, the number of $\delta$-sunspots are simply not a fraction of the strength or total sunspot number of the SC. One would have to wait
till the end of SC 25 to determine the total number of $\delta$-ARs which emerge and verify if the above statement holds true or not.      

The $\delta$- and non-$\delta$-phases have median lifetimes of around five and eight days, respectively and 75\% of ARs in both phases 
produce less than 20 flares. 
However, in terms of the maximum number of flares, the $\delta$-phase is more flare productive, wherein 
80 events and above occur for lifetimes greater than six days reaching peak values of 156, while the maximum number of flares 
produced by the latter is just 68. It should be noted that of the 7651 flares attributed to the non-$\delta$-phase across all three SCs, 
about 60\% of the flares are associated with a $\beta\gamma$-configuration which along with the $\delta$-configuration account for nearly 80\% 
of all flares originating from a $\delta$-AR. While the $\delta$-phase does produce more flares, Fig.~\ref{fig06} shows that about 60--78\% 
of the cases are flare productive with lifetimes between 6--12 days. In comparison, the non-$\delta$-phase, for the same lifetimes, has a 
higher proportion of flare productive cases ranging from 72--96\%. However, it is also observed that in the $\delta$-phase, the majority 
of ARs that did not produce any flares, having lifetimes of 3--13 days, were recorded prior to 2003 and account for about 70\% and higher of such 
cases. If one assumes that the lack of flares in the database prior to 2003 was the result of observational biases or lack of reporting, and 
the fraction of ARs producing flares during the $\delta$-phase were similar to the non-$\delta$-phase, there still remains a fraction 
of 10--20\% of ARs that possess a $\delta$-type magnetic configuration but do not produce any flares. This aspect has been pointed out by 
\cite{1986SoPh..103..111P} wherein additional conditions of strong horizontal gradients in the longitudinal and transverse magnetic fields 
could decide whether a $\delta$-sunspot would produce flares or not. \cite{2014A&A...562L...6B} found that a small $\delta$-sunspot could 
go as long as 24 hr without producing any flares. It has also been shown that the flare productivity of $\delta$-sunspots also varies with 
the epoch of the SC and can be different for different cycles \citep{2020ApJ...894...77G}. 

The latitudinal distribution of $\delta$-ARs across the northern and southern hemispheres is nearly symmetric on either
side of the equator for SCs 23 and 24, peaking around $\pm$10$^\circ$--20$^\circ$, which is consistent with the 
results of \cite{2019A&A...629A..45N}. On the other hand, the latitudinal distribution in SC 25 tends to be asymmetric and weaker possibly 
due to the cycle still being in progress. The maximum area of the host AR follows the peak of the latitudinal distribution that is dominated by
the southern hemisphere, particularly in SCs 24 and 25. However, the extent of the symmetry is possibly dictated and dominated by the fraction 
of $\delta$-ARs, whose area exceeds the mean area at each latitudinal bin, that is around 30\% of the sample. This characteristic is observed 
in SCs 23 and 24 but not in SC 25, again possibly due to the cycle being incomplete.  

The final point that I would like to raise is concerning data availability and the methods/techniques used at different facilities to derive 
quantities from the observations. The sunspot data utilized in this work, particularly the NOAA AR number, Hale classification, and sunspot area, 
were taken from NOAA's SWPC which were compared with the records from SIDC, that reveal data gaps in the former. In addition, the data 
processing routines, which are employed to extract features and quantities, have intrinsic uncertainties and limitations, which could 
introduce differences between various resources. As shown by \cite{2016SoPh..291...41P}, incorporating automatic schemes for the 
detection of $\delta$-sunspots yields noticeable differences with manual identification, where both under- and over-estimation can be seen.
The excess number of $\delta$-ARs between 2001--2002, as seen in Fig.~\ref{fig02}, is possibly an artifact arising at the 
classification stage at SWPC/SolarMonitor. The result of the over-counting makes it difficult to study temporal variations of quantities such as
the Hale class, AR area, number of spots, etc. within a single solar cycle and the only alternative would be to look at the net cyclic 
behavior of these quantities from one cycle to the other. One possible way to circumvent this would require an independent detection scheme 
combining both automated and human confirmation. The above issue is also valid for flare data from the 
various online resources which, although use the same source of data, may introduce errors in the entries which can be difficult to 
identify and sort, automatically. For instance, NOAA AR 13912 produced an X2.3 flare at 08:50 UT on 2024 December 08, according to Solar Monitor. 
The same event in the HEC appears as an X2.2 flare with all other details remaining the same. Instruments on board Hinode and 
the Solar Dynamics Observatory \citep[SDO;][]{2012SoPh..275....3P} have been observing the Sun for around 18 and 15 years, respectively under stable conditions
for this length of time. Any changes 
in the feature identification and processing scheme would have at least affected solar cycles 24 and 25 in the same manner and not in one specific 
cycle over the other. In order to extend the information over several solar cycles it is necessary to cross-calibrate and verify data from 
various instruments and facilities that overlap in their operational lifetimes and require supervision by the international community using a standardized 
procedure to ensure there are no artifacts or biases in the catalogs. Thus, one should exercise caution, when 
describing trends, if a universal data processing scheme is not employed or systematic differences from various data resources are not reconciled 
periodically, that could affect the interpretation of the results and the conclusions drawn from them.

\section{Conclusions}
\label{conclude}
Sunspots or active regions with a $\delta$-magnetic configuration are known to be associated with strong eruptions such as 
flares and mass ejections. 
This article, investigates the relationship between $\delta$-active regions and flares over the course of 
three solar cycles, from 1996 to 2024, with respect to the former's area, lifetime, latitudinal distribution, and the phase of its 
magnetic complexity. 
Solar cycle 25, while still in progress, has produced the least number of $\delta$-active regions in comparison 
to the previous two solar cycles, yet the number of M- and X-class flares exceed that of cycle 24 by 25\%. Flare occurrence is 
higher in C-, M-, and X-class events during the presence of the $\delta$-configuration in an AR, which is seen in all 
three solar cycles. The total number of flares produced by $\delta$- and  non-$\delta$-active regions were 15875 and
17033, respectively across all three solar cycles. The latter are dominated by B- and C-class flares, while the number of M- and X-class 
flares across all three SCs was significantly less than compared to $\delta$-ARs.
The median lifetime of an active region in the $\delta$-phase is about five days while it is about eight days in the non-$\delta$-phase. 
The typical number of flares produced by a $\delta$-active region is 20, however, about 30\% of $\delta$-active regions do not produce 
flares when their lifetimes are between 6--12 days. The latitudinal distribution of $\delta$-active regions across the northern and 
southern hemispheres is nearly symmetric on either side of the equator for solar cycles 23 and 24, peaking around $\pm$10$^\circ$--20$^\circ$. 
For solar cycles 23 and 24, about 30\% of the $\delta$-active regions have an area exceeding the mean value over the above latitudinal 
belt while for solar cycle 25, there is a large scatter possibly due to the cycle still being in progress. It remains to be seen if the 
latter phase of solar cycle 25 will be as active as its earlier phase and whether the number of $\delta$-active regions emerging during that 
period scale with the total sunspot number.

\begin{figure} 
\centerline{
\includegraphics[angle = 90,width=\textwidth]{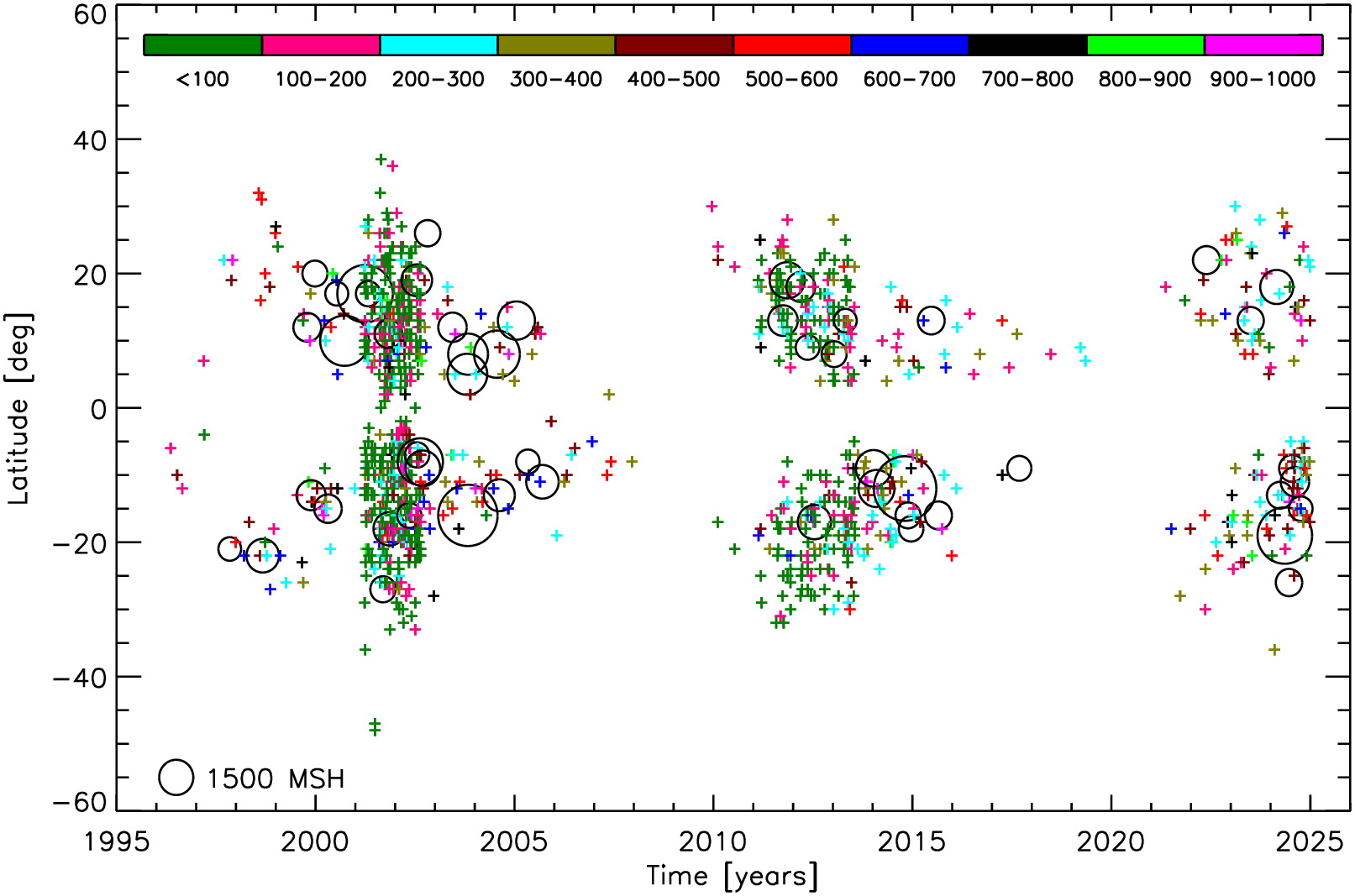}
}
\vspace{-5pt}
\caption{Butterfly diagram of $\delta$-ARs on the Sun from during SC 23 to SC 25. The colored plus symbols correspond to the area of the AR 
in MSH as indicated in the color bar on the top of the panel, while the circles depict $\delta$-ARs with an area greater than 1000\,MSH with 
the radius representing the area normalized by a value of 3500\,MSH.}
\label{app-fig01}
\end{figure}

%
\appendix   
\label{appendix}
Figure~\ref{app-fig01} shows the butterfly diagram of $\delta$-ARs on the solar disc over the three SCs, respectively.
The plus symbols and circles refer to the time and latitude midway of an AR's transit when it comprised a 
$\delta$-magnetic configuration. The plus symbols correspond to $\delta$-ARs having an area less than 1000\,MSH with the colors
representing the area as indicated in the color bar on top of the panel. The circles correspond to $\delta$-ARs having an area greater than
1000\,MSH with the radius indicating the maximum area of the $\delta$-AR during its transit. 

%
\begin{acks}
\href{https://solarmonitor.org/index.php}{SolarMonitor.org}
is maintained and developed by a team of researchers in the School of Physics, Trinity College Dublin and
is funded by ESA/PRODEX and a grant from the EC Framework 7 Programme (HELIO).
The HEC is part of the Heliophysics Integrated Observatory, HELIO, a Research Infrastructure that 
addresses the needs of a broad community of researchers in Heliophysics and was funded under the Capacities 
Specific Programme within the European Commission's Seventh Framework Programme (FP7; Grant No. 238969). 
Hinode is a Japanese mission developed and launched by ISAS/JAXA, with NAOJ as domestic partner and 
NASA and STFC (UK) as international partners. It is operated by these agencies in cooperation with 
ESA and the NSC (Norway). This catalog was developed by the Smithsonian Astrophysical Observatory 
and is based on the \href{https://hinode.isee.nagoya-u.ac.jp/flare_catalogue/}{Hinode Flare Catalog} 
maintained by ISAS/JAXA and the Institute for Space-Earth Environmental Research (ISEE) at Nagoya University.
REL wishes to thank the anonymous referee for reading the manuscript meticulously and providing insightful comments.
\end{acks}

 \begin{ethics}
 \begin{conflict}
 The author declares no competing interests.
 \end{conflict}
 \end{ethics}

%

\bibliographystyle{spr-mp-sola}
\bibliography{louis_ref}

\end{document}